\documentclass{aastex6}

\usepackage{graphicx}

\begin{document}

\title{Discovery of Tidal RR LYRAE stars in the Bulge Globular Cluster M 62\footnote{Based on observations collected at the European Organisation for Astronomical Research in the Southern Hemisphere under ESO programmes 179.B-2002 and 298.D-5048.}	
}

\author{
Dante Minniti\altaffilmark{1,2,3},
Jos\'e G. Fern\'andez-Trincado\altaffilmark{4,5,6},
Vincenzo Ripepi\altaffilmark{7},
Javier Alonso-Garc\'ia\altaffilmark{8,2},
Rodrigo Contreras Ramos\altaffilmark{9,2},
Marcella Marconi\altaffilmark{7}
}

\altaffiltext{1}{Depto. de Cs. F\'isicas, Facultad de Ciencias Exactas, Universidad Andr\'es Bello, Av. Fern\'andez Concha 700, Las Condes, Santiago, Chile.}
\altaffiltext{2}{Millennium Institute of Astrophysics, Av. Vicuna Mackenna 4860, 782-0436, Santiago, Chile.}
\altaffiltext{3}{Vatican Observatory, V00120 Vatican City State, Italy.}
\altaffiltext{4}{Instituto de Astronom\'ia y Ciencias Planetarias, Universidad de Atacama, Copayapu 485, Copiap\'o, Chile.}
\altaffiltext{5}{Institut Utinam, CNRS-UMR 6213, Universit\'e Bourgogne-Franche-Compt\'e, OSU THETA Franche-Compt\'e, Observatoire de Besan\c{c}on, BP 1615, 251010 Besan\c{c}on Cedex, France.}
\altaffiltext{6}{Departamento de Astronom\'ia, Casilla 160-C, Universidad de Concepci\'on, Concepci\'on, Chile.}
\altaffiltext{7}{INAF-Osservatorio Astronomico di Capodimonte, via Moiariello 16, 80131, Naples, Italy.}
\altaffiltext{8}{Centro de Astronom\'{i}a (CITEVA), Universidad de Antofagasta, Av. Angamos 601, Antofagasta, Chile.}
\altaffiltext{9}{Instituto de Astrof\'isica, P. Universidad Cat\'olica, Av. Vicuna Mackenna 4860, 782-0436 Macul, Santiago, Chile.}

\begin{abstract}
The RR Lyrae (RRL) rich globular cluster M 62 (NGC 6266) is one of the most massive globular clusters in the Milky Way, located in the dense region of the Galactic bulge, where dynamical processes that affect the survival of globular clusters are maximised. Using Gaia DR2 data we have found clear evidence for an excess of RRLs beyond the cluster tidal radius of M 62, associated partly with stars stripped into the Galaxy field. This is confirmed with new VVVX survey observations, that discard any differential reddening effect as the possible cause of the observed RRL density excess.  We also determined the orbit of M 62 using accurate new measurements of its distance, radial velocity and proper motions (PMs), finding that its orbit is prograde respect to the direction of the Galactic rotation. Orbits are integrated in the non-axisymmetric galactic model \texttt{GravPot16}, which includes the perturbations due to the central Galactic bar. M 62 shows a particular orbital behaviour, having a dynamical signature of the bar-bulge region. The small extra-tidal RRLs extensions observed are roughly aligned towards the galactic center and the direction almost perpendicular to the galactic plane and not with its motion along its orbit, which may be a clear sign of bulge-crossing shocks, during the last passage close of the cluster towards its perigalacticon. M 62 would be the first clear observed case of bulge shocking in the inner Galaxy acting on a globular cluster.
\end{abstract}

\section{Introduction} 
\label{sec:intro}

Globular clusters that we observe today in the Milky Way are the survivors of a much larger initial population that was decimated by dynamical processes, chiefs among them being  tidal distruption and  dynamical friction (Fall \& Rees 1977, 1985).
Tidal disruption of globular clusters has been widely discussed (e.g. King 1962, Tremaine et al. 1975, Chernoff et al. 1986, Capuzzo-Dolcetta 1993, Weinberg 1994, Meylan \& Heggie 1997, Gnedin \& Ostriker 1997, Vesperini \& Heggie 1997, Combes et al. 1999, Lotz et al. 2001, Capuzzo-Dolcetta et al. 2005, Balbinot \& Gieles 2017), and prominent examples of tidal tails have been observed  (e.g. Leon, et al. 2000, Odenkirchen et al. 2001, Belokurov et al. 2006, Grillmair et al. 2006, Jordi \& Grebel 2010). 
In this paper we report the discovery of a potential extra-tidal RRLs population, likely the results, in projection on the plane sky, of the relics of the bulge shocking.

M 62 is one of the most massive globular clusters in the Milky Way, and also one of the richest clusters in RRL variable stars (Contreras Ramos et al. 2010). This cluster has an extended blue horizontal branch (BHB), and a strong EUV excess observed by IUE (Dorman et al. 1995), and dynamically it does not appear to have suffered core collapse yet (Beccari et al. 2006). 
It is located in the Galactic bulge at $l, ~b = (353.57458, ~+07.31956)$ deg, at a distance of D=6.9 kpc (Harris 1996), and its orbit confines the cluster to the bulge (Dinescu et al. 2003). 

This cluster is  interesting also because it has a double main-sequence, indicative of a composite stellar population (Milone et al. 2015), like $\omega$ Cen, the largest Galactic globular cluster, that has been proposed to be the remaining nucleus of an accreted dwarf galaxy.

In addition, this cluster shows evidence for overdensity features beyond the tidal radius (Han et al. 2017). However, the cluster is located in a field with significant differential reddening (Minniti, Coyne \& Claria 1992, Contreras Ramos et al. 2005), that severely diminishes the capacity to recognise extra-tidal stars.

Because of its known physical properties (location, mass, radius, 3-D velocity), M 62 is an ideal laboratory to explore the effects of the different dynamical processes that affect the survival of massive clusters in the inner regions of the Milky Way. 

In particular, in this work we discuss the effects of tidal disruption on M 62 using RRLs in the Gaia DR2 database (Gaia Collaboration et al. 2016, 2018, Clementini et al. 2018), and new near-IR observations of the VISTA Variables in the Via Lactea Extended public survey (VVVX, Minniti et al. 2010, Saito et al. 2012).
The latter are important to complement the existing optical observations that make difficult the search and interpretation of the possible tidal tails in the presence of differential reddening.

This Letter is organized as follows.
In \autoref{sec:sec2} we describe the data used for the discovery of extra-tidal RRL stars in M 62.
In \autoref{sec:sec3} we present the M 62 orbital models.
In \autoref{sec:sec4} we discuss the two dominant dynamical processes, dynamical friction vs tidal disruption.
Finally, our conclusions are summarized in \autoref{sec:sec5}. \\

%%%%%%%%%%%%%%%%%%%%%%%%%%%%%%%%%%%%%%%%%%%%%%%%%%%%%%%%%%%%%%%%%%%%%%%%%%%%%%%%
\section{Discovery of Extra Tidal RRL in M 62}
\label{sec:sec2}

\begin{figure}[t]
\begin{center}
\includegraphics[height = 9 cm]{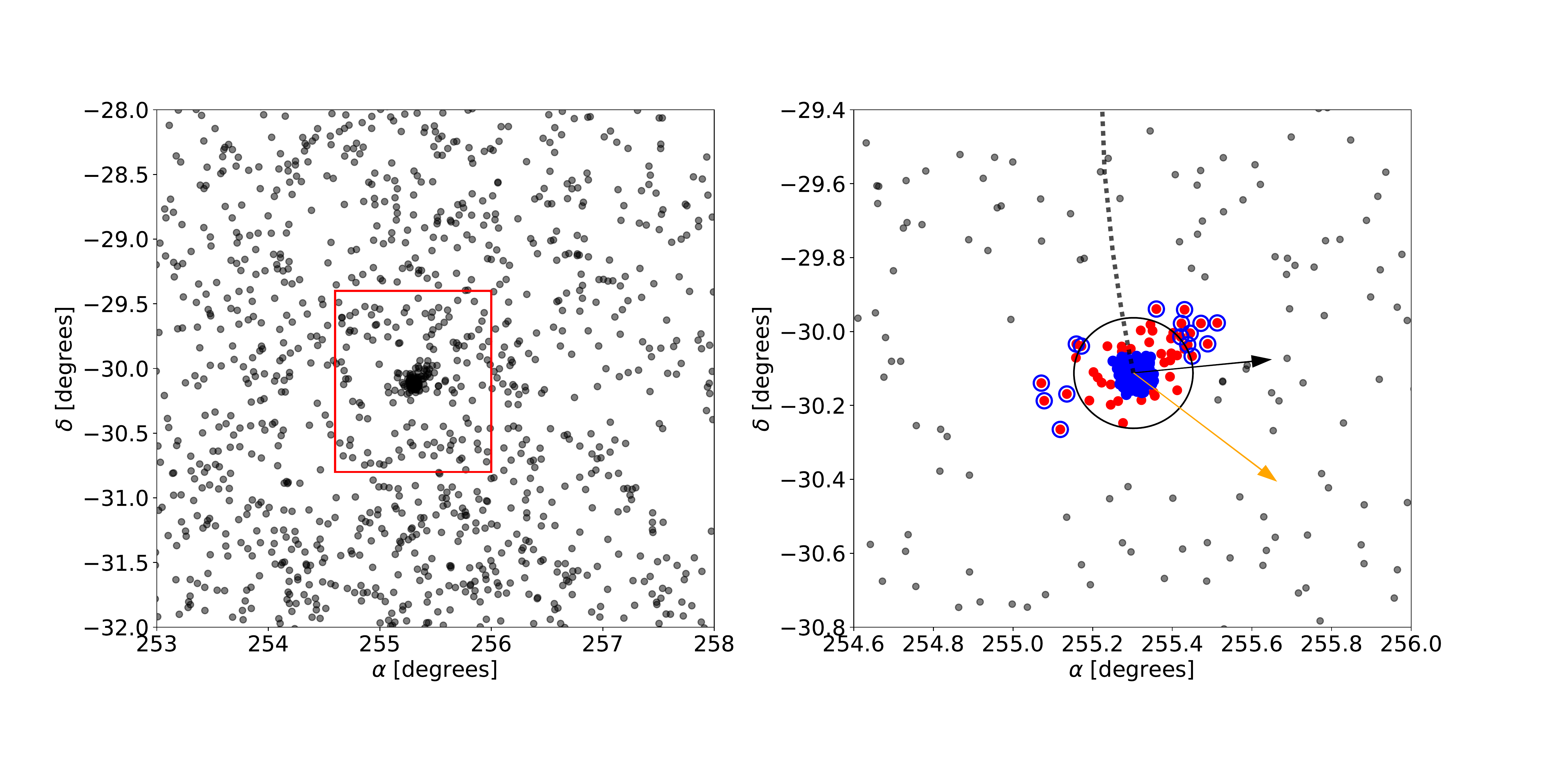}\\
\caption{  
Left: Large scale distribution of RRLs (black dots) in the vicinity of the globular cluster M 62. 
Right: Zoom in the M 62 region (indicated by a red box in the left panel), showing individual RRL stars in our sample, with the cluster tidal radius over-plotted. Most of the cluster RRLs are located within $5 r_h \approx 5$ arcmin of the cluster center (blue dots). The filled red circles designate RRLs with $K_{s}< 14.5$, while the blue open circles designate the RRLs outside the cluster tidal radius. The black and orange arrows point to the Galactic center and to the Galactic plane, respectively. The black dotted line curve is the orbital path of the cluster.}
\label{fig1}
\end{center}
\end{figure}

Recently Clementini et al. (2018) presented the first all-sky map of RRL  and Cepheid variable stars observed by the
Gaia mission during the first 22 months of science operations, including new RRL discoveries.  We explored these Gaia second data release (DR2)  RRLs in the region surrounding the globular cluster M 62.
The  density distribution of RRLs in the vicinity of this globular cluster is presented in Figure \ref{fig1}, which shows a clear and small extension of extra-tidal candidates extending from the cluster in the South-West to North-East direction,
 beyond the tidal radius of the cluster, ($R_t=8.97$ arcmin, Harris 1996), and aligned $\sim$ 74 degrees away from the predicted orbital path of M 62 and roughly aligned ($\sim$ 19 degrees) towards the galactic center direction.  It is very likely that the tidal tails visible in Figure \ref{fig1} are the results of bulge shocking and likely related to its strong interaction with the Galactic bar.
 
Using the Gaia RRL catalog of Clementini et al. (2018), we measure a background density of bulge RRLs of $N=71 \pm 6$ RRL/sq. deg. in a 2 sq. deg. field surrounding the cluster. We find an excess of a dozen RRL beyond the tidal radius of M 62, in a region where fewer than 3 RRLs were expected. Arguably this is only a $\simeq 3 \sigma$ result with small number statistics, but the asymmetry is even more marked if we consider stars inside the tidal radius. For example, considering stars outside 2/3 $r_t = 6 r_h$ from the cluster center, we count 8 RRLs in the South-West direction of the cluster and 25 in the North-East direction of the cluster (vs 5 expected field RRL for the same area). 

The total number of M 62 RRLs is $N=209$ (Contreras Ramos et al. 2010), most of which are located inside $5 r_h\approx 5$ arcmin of the cluster center (Figure \ref{fig1}). Therefore, the RRL excess is significant, of the order of $\sim 10\%$ of the total, and
this is probably a lower limit, because of the projection effect of the cluster orbital inclination. 
Apparently, if all of these are RRL extracted from the cluster, this is a significant effect, as it would imply that the future survival of M 62 is restricted to a few Gyrs. However, careful modelling is needed in order to predict the future orbital decay in the presence of the cluster mass loss.

\begin{figure*}[t]
\begin{center}
\includegraphics[height = 10 cm]{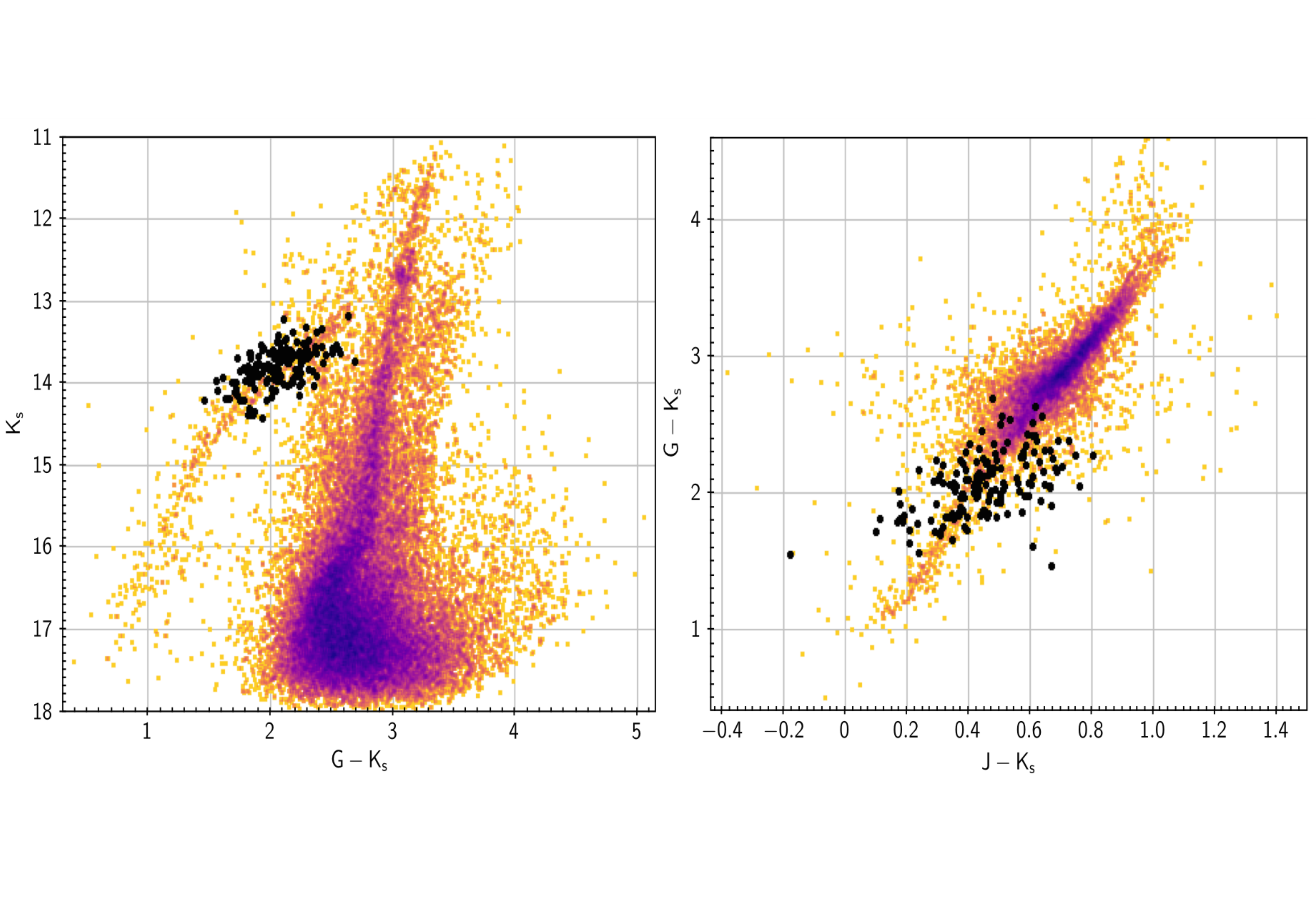}\\
\caption{
Left:{Optical-near-IR color-magnitude diagram of M 62 from the VVVX survey, after correcting from differential reddening using the extinction maps in Alonso-Garc\'ia et al. (2012). The Gaia RRL with VVVX photometry are overplotted as large black circles.}
Right: Optical-near-IR color-color diagram of M 62 from the VVVX survey, overplotted with the Gaia RRL  (large black circles). 
The observed RRL scatter in these diagrams is due to the fact that the VVVX near-IR photometry comes from single epoch observations, and that their photometry is not  corrected for differential reddening.
}
\label{fig2}
\end{center}
\end{figure*}

The field of M 62 exhibits significant differential reddening (e.g. Minniti et al. 1992, Contreras Ramos et al. 2005, Alonso-Garcia et al. 2012), and the variable extinction may severely diminish the capacity to recognise extra-tidal stars.
In order to check for the effects of differential reddening we inspect the near-IR photometry of the VVVX survey. Figure \ref{fig2} shows the reddening corrected optical-near-IR color-magnitude and color-color diagram for this cluster, covering a region inside its tidal radius. Overplotted are the Gaia RRL stars in the field, discarding the background RRL with $Ks>14.5$. 
It is important to determine that this RRL excess is not a chance alignment of stars along the line of sight. Because RRL are primary distance indicators, they are optimal tracers as the photometry can confirm that they are located at the distance of M 62, $D=6.9$ kpc (Harris 1996).

RRL variable stars show periodic variations in their magnitudes and colors, which account for the observed scatter, because the color-magnitude and the color-color diagrams are based on single-epoch near-IR photometry. Figure \ref{fig2} also shows that the influence of reddening is not significant in this RRL sample.

\begin{figure}[t]
\begin{center}
\includegraphics[height = 8 cm]{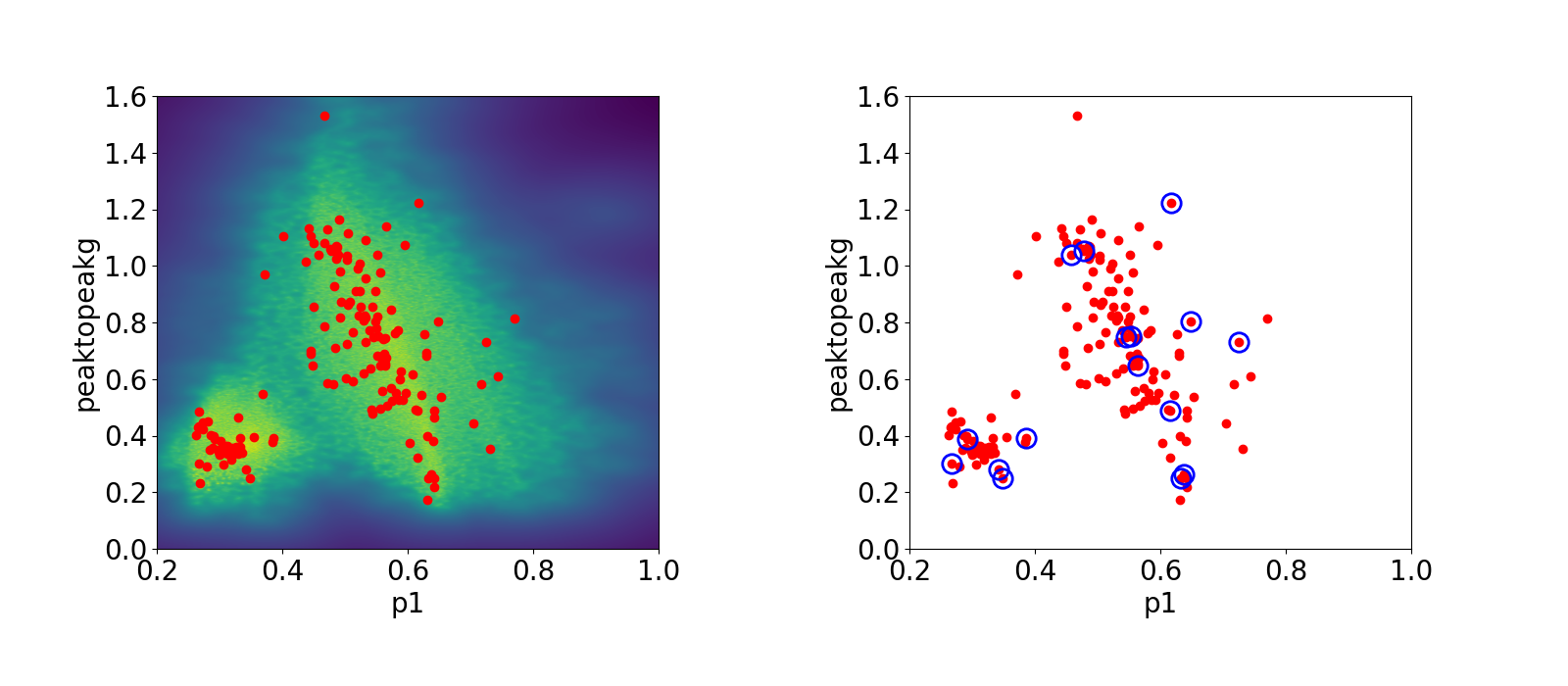}\\
\caption{
{Left:  Kernel Density Estimate (KDE) smoothed distributions of the Amplitude vs. period for field RRL stars from Gaia DR2 data (see Clementini et al. 2018) with the M 62 RRLs (red dots) in our sample over-plotted}. 
{Right: Comparison of the Bailey diagram for stars in the main body of M 62 (red dots) with the RRL located outside the tidal radius of the cluster (blue open circles).}
}
\label{fig3}
\end{center}
\end{figure}

We have to characterize the extra-tidal  RRL in M 62 in order to check if they belong mostly to the field or to the cluster, or to a combination of both.
In the case of tidal disruption, we expect that they belong to the cluster. In order to characterise the extra-tidal RRL, we compare them with the field and cluster RRLs using the color-magnitude, the color-color, and the period-amplitude (Bailey) diagrams (Figure \ref{fig2} and \ref{fig3}). 
These comparisons, however, are not conclusive, as it is clear from these diagrams that the parameters of the M 62 RRLs are similar to those of the field stars. 
This situation would  improve and a better discrimination between cluster and field stars will be possible when the final Gaia data release delivers more accurate individual RRL PMs in a few years.

\section{The Orbit of M 62}
\label{sec:sec3}

Having found a clear excess of extra-tidal RRL in M 62, it is important to determine the cluster orbit in order to seek for possible explanations for this feature.
Casetti et al. (2003) measured the PMs and computed the orbit for this cluster, finding a prograde orbit confined to the bulge. 
In this section we recompute the orbit, using the new Gaia PM data, and the novel Galaxy modelling algorithm \texttt{GravPot16}
{\footnote{https://Fern\'andez-trincado.github.io/GravPot16/}} (Fern\'andez-Trincado et al., in preparation).

\begin{figure}[t]
\begin{center}
\includegraphics[height = 18 cm]{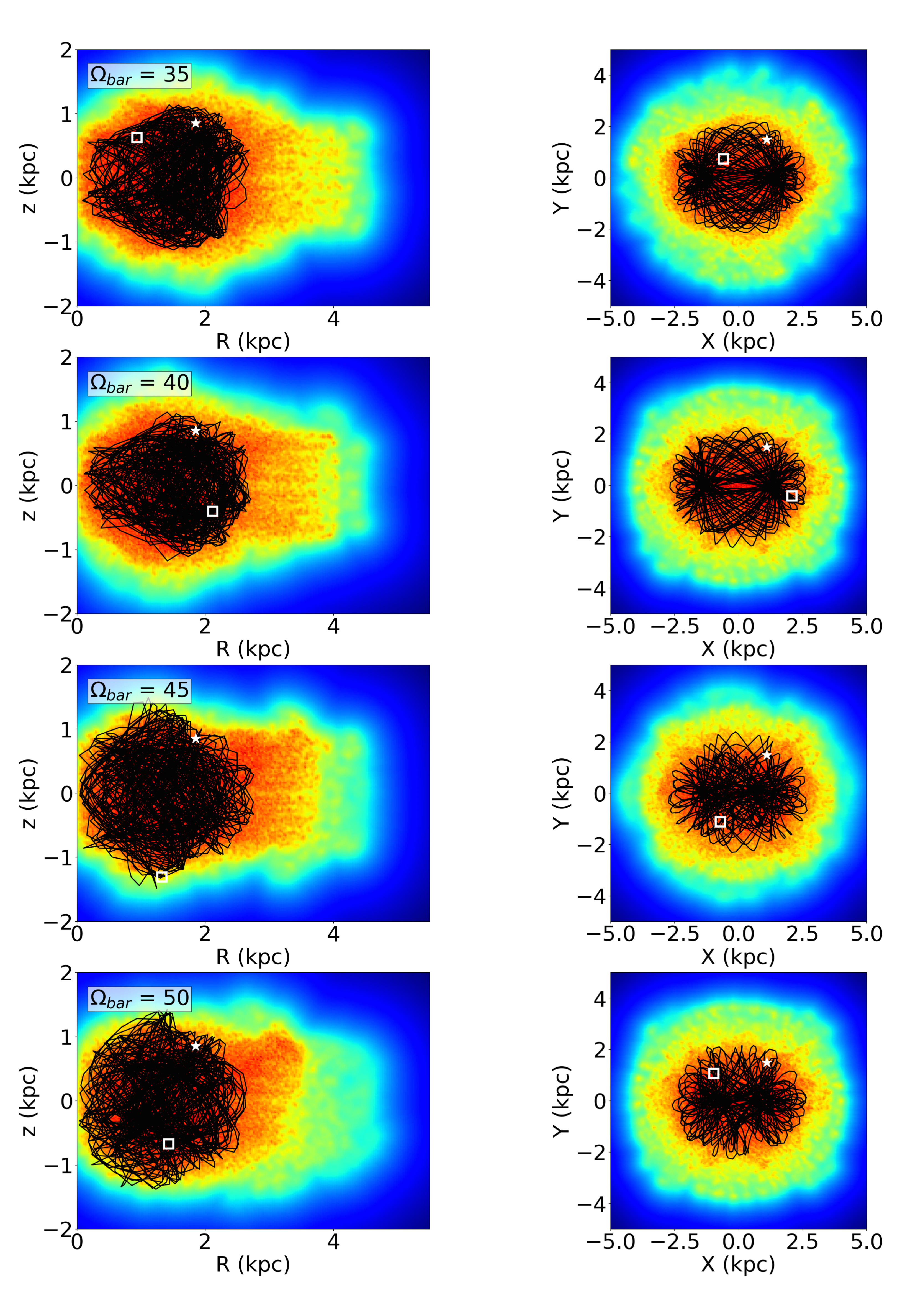}\\
\caption{
Orbital simulations for the globular cluster M 62 in the z vs $R_G$  (left) and y vs x (right) planes.
The small white star marks the present position of the cluster, and the white square marks its final position.
The solid black lines show the central orbital values, while the colored regions show the wider range of allowed orbits taking into account the errors in the initial parameters.
}
\label{fig4}
\end{center}
\end{figure}

\begin{figure}[t]
\begin{center}
\includegraphics[height = 16 cm]{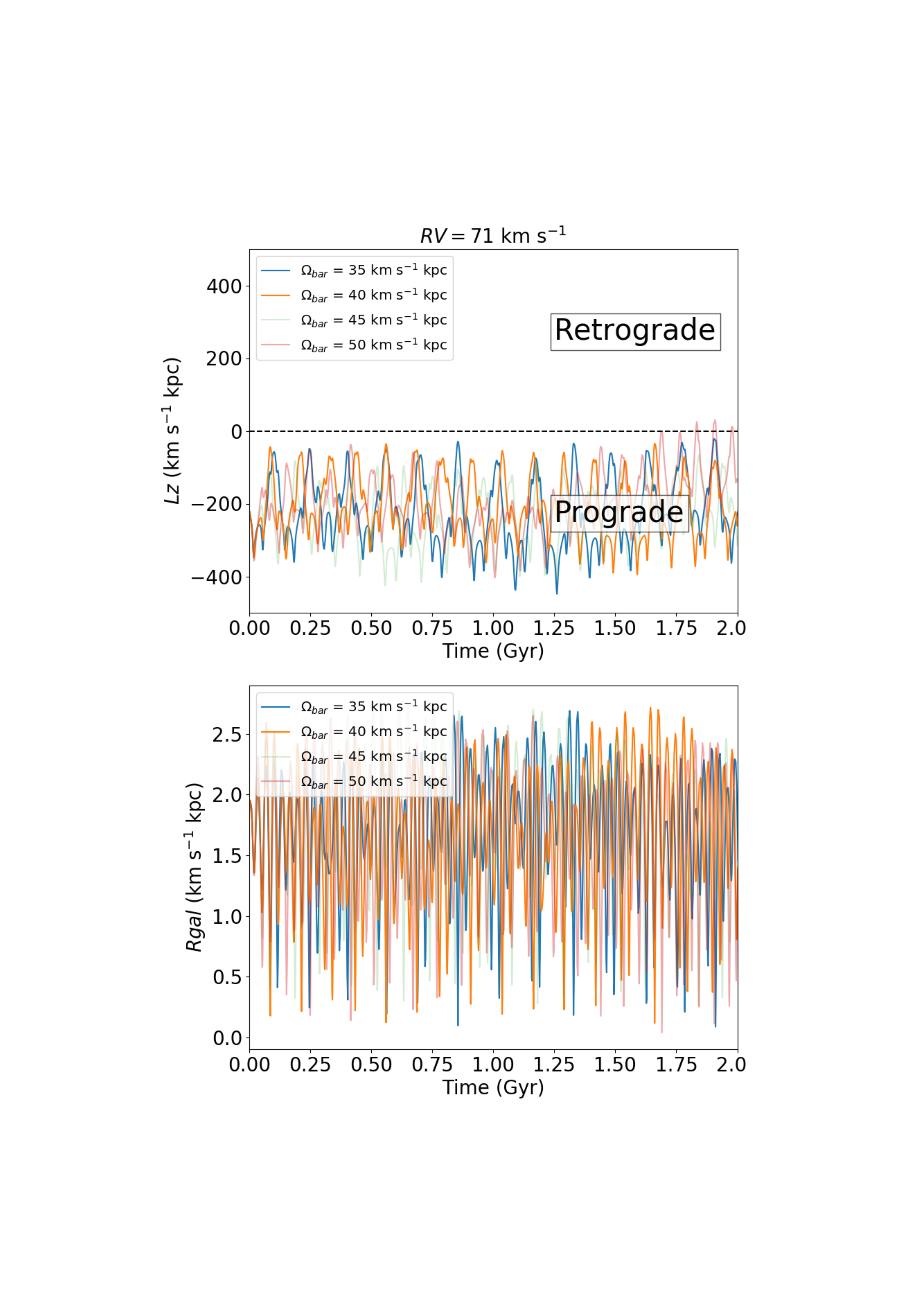}\\
\caption{
Top. Time evolution of the vertical component of the angular momentum $L_z$ in km/s/kpc for the M 62 orbital integrations, showing that the orbit remains prograde at all times.
Bottom. Time evolution of the M 62 Galatocentric distance $R_G$  in kpc for the orbital simulations with different bar pattern speeds, as labelled. The cluster gets very clese to the Galactic center ($R_G~0.2$ kpc) periodically.
}
\label{fig5}
\end{center}
\end{figure}

We have integrated the orbits following similar procedures that we have applied elsewhere (Fern\'andez-Trincado 2016b,  Contreras Ramos et al. 2018), and using four different bar pattern speed values ($\Omega_{bar}=35, 40, 45, 50$ km/s), and the same dynamical model configuration as described in Contreras Ramos et al. (2018). 
Figure \ref{fig4} shows the central values of the orbits, revealing that M 62 is a globular cluster that is trapped into the Galactic bar structure, and in resonance, as shown by the boxy orbital structure in the X vs Y plane. 
Figure \ref{fig5} shows the time evolution of the vertical component of the angular momentum $L_z$ in km/s/kpc, where negative $L_z$ in our reference system means that the cluster orbit is prograde (in the same sense as the disk rotation), confirming that the cluster is dynamically trapped into the bar potential. We have also computed Monte Carlo simulations using the range of input uncertainties, for two different sets of initial configurations (radial velocities and PMs).
For the orbital results shown in Figure \ref{fig4}, we have used the following initial inputs:
$\mu_{RA}  = -4.96 \pm 0.01$ mas/yr,
$\mu_{DEC} = -2.96 \pm 0.01$ mas/yr, 
$V_r = -71.0  \pm  0.7$ km/s and
$D=6.9$ kpc, estimated from 53 selected cluster members\footnote{ A 2-dimensional Gaussian smoothing routine was applied in proper motion space for stars with g $<$ 17 mag. Two samples are computed: all stars within twice the cluster radius and stars outside the cluster's radius, then the outside sample is subtracted from the full sample. A 2D Gaussian is fit to the remaining peak and membership probabilities are assigned.} with available PMs and radial velocity from Gaia data.
We have also run the same simulations using slightly different inputs taken from \citep{Helmi2018}, e.g.: 
$\mu_{RA} = -5.33 \pm 0.01$ mas/yr,
$\mu_{DEC} = -2.98 \pm 0.01$ mas/yr, and
$V_r  = -74.9 \pm 0.8$ km/s,
but find that the results do not change at all. In addition to overall agreement between the inputs adopted in this work and Helmi's values, our results also agree remarkably well with the more recent measurements of PMs and radial velocity for M 62, e.g.: $\mu_{RA} = -5.06 \pm 0.01$ mas/yr, and $\mu_{DEC} = -3.02 \pm 0.01$ mas/yr from Vasiliev (2018). We note that the computed orbital elements are similars even adopting the recent available inputs of M 62, and therefore do not affect the results presented in this work.

The Monte Carlo solutions confirm that the cluster is trapped in the structure of the bar, independently of the adopted bar pattern speeds. Changing the bar pattern speed changes the resonance orbital family  (Figure \ref{fig4}), but the cluster remains trapped into the bar potential.
We added a white star symbol to indicate the present custer position, and a square to show the end position after the integrations. 

Figure \ref{fig5}  also shows the variation of the cluster Galactocentric distance $R_G$ with time, starting from the initial position to a fraction of the orbit integration time. 
The cluster has recently passed through the apogalacticon, the most distant point along its orbit, and is now heading for perigalacticon.
Although this cluster is presently located at $R_G = 1.7$ kpc, it can get as close as $R_{peri}=0.2$ kpc from the Galactic centre during its orbital motion. 
The cluster tidal disruption and mass loss would be critical during these perigalactic passages. 
However, we have not considered mass loss that may be considerable and  would produce orbital delay for the cluster.
 A full exploration of M 62 models including the possible effects of dynamical friction and tidal evaporation is warranted.

\section{DISCUSSION}
\label{sec:sec4}

The small extra-tidal extension of RRLs observed beyond the tidal radius of M 62 is roughly ($\sim$19 degrees) aligned towards the galactic center and the direction almost perpendicular to the galactic plane and not with its motion along its orbit. Given its position in the Galaxy, this cluster is probably suffering a strong bulge shocking, with the tail almost aligned towards the tidal directions after being compressed during the crossing by its perigalacticon. They are therefore much less likely to be field stars associated to the so-called gravitational wake effect, that exerts a drag force that slows down the cluster motion (Chandrashekhar 1943, Tremaine \& Weinberg 1984, Capuzzo-Dolcetta \& Vicari 2005, Arca-Sedda \& Capuzzo-Dolcetta 2014). 
	
In the particular case of M 62, we see a concentration of stars aligned $\sim$74 degrees away from the predicted orbital path of M 62 (Figure \ref{fig1}), as expected of tidally stripped cluster stars (e.g., Fern\'andez-Trincado et al. 2015a, 2016a), with a huge extension in the north-east part, possibly attributed to that they do not all escape from a single Lagrange point on the edge of the cluster. 

Weighting all that evidence, we conclude that the extra-tidal excess of RRL seen in Figure \ref{fig1} is indeed signatures of tidal debris around M 62 caused by  tidally stripped stars in the bulge region and it is very likely that the tidal tails of M 62 are the relics of the bulge shocking from the last passage close to the galactic center.

This globular cluster therefore is a prime target to compute detailed dynamical models, in order to estimate its survival time before it is dragged into the innermost Galactic region. This finding support the idea of the former globular cluster member stars have been indeed recently found in the inner bulge (Schiavon et al. 2017b; Fern\'andez-Trincado et al. 2017b), suggesting that tidal effects in the Galactic bulge could have been efficient over time.

In addition, using future measurements of individual radial velocities  in combination with the PMs measured by Gaia, it would be possible to compute the dynamics of the individual stars surrounding this cluster. 
Unfortunately, the recent M 62 radial velocity measurements  of Kamann et al.(2017) are all located inside the cluster tidal radius. 
Also, it would be important to map different stellar tracers (such as red clump giants, blue horizontal branch stars) in a large field around M 62 in order to search for more extended tidal tails. This would be a daunting task, owing to the differential reddening and high stellar field density.
 
\section{Conclusions}
\label{sec:sec5}
The RRL rich globular cluster M 62 is one of the most massive globular clusters in the Milky Way. It is located in the dense bulge region, deep in the Galactic potential well where dynamical processes that affect the survival of globular clusters are maximised.

Using Gaia DR2 RRL from Clementini et al. (2018), we have found clear evidence for an excess of stars beyond its tidal radius. This is confirmed with new VVVX observations, that discard any differential reddening effects as the possible cause of the observed stellar density excess. This extra-tidal excess can be interpreted either as the detection of the cluster "tidal tails", or as dissolving cluster stars due to tidal forces inside the bulge, and particularly evidence of bulge shocking.  
 
We determined the orbit of M 62 using accurate new measurements of its distance, radial velocity and PMs.
We find that its orbit is prograde and confined to the Galactic bulge, in agreement with the earlier results by Dinescu et al. (2003). In addition, we find that M 62 is probably trapped in a resonance orbit within the Galactic bar. This orbital configuration within the dense bulge region, along with the cluster mass and velocity would maximise the effects of bulge shocking.

We weighted the observational evidence against the possible dominant dynamical process that affect the survival of massive clusters in the inner regions of the Milky Way: tidal disruption. While the more feasible interpretation of the observed RRL excess as possible tidal tails cannot be rule out, we argue that tidal disruption due to bulge shocking not too far from the Galaxy center sis the best explanation in this case, because of the observed asymmetry in the RRL distribution that is opposite with the cluster motion and the tails are elongated mainly along the galactic density gradient.

\acknowledgments
This work has made use of data from the European Space Agency (ESA) mission {\it Gaia} (\url{https://www.cosmos.esa.int/gaia}), processed by the {\it Gaia}
Data Processing and Analysis Consortium (DPAC, \url{https://www.cosmos.esa.int/web/gaia/dpac/consortium}). Funding for the DPAC has been provided by national institutions, in particular the institutions
participating in the {\it Gaia} Multilateral Agreement. We thank the support by Istituto Nazionale di Astrofisica (INAF) and the Agenzia Spaziale Italiana (ASI) through grants I/037/08/0, I/058/10/0, 2014-025-R.0, and 2014-025-R.1.2015 to INAF (PI M.G. Lattanzi). We also gratefully acknowledge data from the ESO Public Survey program ID 179.B-2002 taken with the VISTA telescope, and products from the Cambridge Astronomical Survey Unit (CASU). Support is provided by the BASAL Center for Astrophysics and Associated Technologies (CATA) through grant PFB-06, and the Ministry for the Economy, Development and Tourism, Programa Iniciativa Cientifica Milenio grant IC120009, awarded to the Millennium Institute of Astrophysics (MAS). D.M. acknowledges support from FONDECYT Regular grant No. 1170121, and the kind hospitality of the Osservatorio di Capodimonte/INAF, Italy. 
J.G.F.-T. is supported by FONDECYT N. 3180210. J.A-G. also acknowledges support by FONDECYT Iniciaci\'on 11150916. V.R. and M.M. acknowledge partial support from Premiale 2015, "MITiC" (P.I. B. Garilli). \\

\mbox{}

\end{document}